# Speaker Identification in Shouted Talking Environments Based on Novel Third-Order Hidden Markov Models

Ismail Shahin
Department of Electrical and Computer Engineering
University of Sharjah
Sharjah, United Arab Emirates
E-mail: ismail@sharjah.ac.ae

*Abstract* **- In this work we propose, implement, and evaluate novel models called Third-Order Hidden Markov Models (HMM3s) to enhance low performance of text-independent speaker identification in shouted talking environments. The proposed models have been tested on our collected speech database using Mel-Frequency Cepstral Coefficients (MFCCs). Our results demonstrate that HMM3s significantly improve speaker identification performance in such talking environments by 11.3% and 166.7% compared to second-order hidden Markov models (HMM2s) and first-order hidden Markov models (HMM1s), respectively. The achieved results based on the proposed models are close to those obtained in subjective assessment by human listeners.**

## I. INTRODUCTION

Speaker recognition has two types: speaker identification and speaker verification (authentication). Speaker identification is the process of automatically deciding who is speaking from a set of known speakers. Speaker verification is the process of automatically accepting or rejecting the identity of the claimed speaker. Speaker identification can be used in criminal investigations to determine the suspected persons who uttered the voice captured at the scene of the crime. Speaker identification can also be used in civil cases or for the media. Speaker verification is widely used in security access to services via a telephone, including home shopping, home banking transactions using a telephone network, security control for private information areas, remote access to computers, and many telecommunication services [1]. Based on the text to be spoken, speaker recognition is categorized into text-dependent and text-independent cases. In the text-dependent case, speaker recognition requires the speaker to generate speech for the same text in both training and testing; on the other hand, in the text-independent case, speaker recognition does not depend on the text being spoken.

## II. LITERATURE REVIEW

Many studies in speech recognition area and speaker recognition area focus on speech uttered in neutral talking environments [1], [2], [3], [4] and on speech produced in stressful talking environments [5], [6], [7], [8]. In literature, many studies that focus on the two areas in stressful talking environments study the two areas in shouted talking environments [8], [9], [10], [11], [12], [13], [14].

Some talking environments are designed to simulate speech generated by different speakers under real stressful talking conditions. Hansen, Cummings, and Clements employed Speech Under Simulated and Actual Stress (SUSAS) database in which eight talking conditions are used to simulate speech uttered under real stressful talking conditions and three real talking conditions [5-7]. The eight talking conditions are neutral, loud, soft, angry, fast, slow, clear, and question. The three talking conditions are 50% task, 70% task, and Lombard. Chen used six talking environments to simulate speech under real stressful talking environments [8]. These environments are neutral, fast, loud, Lombard, soft, and shouted. Shouted talking environments are defined as when speakers shout, their intention is to produce a very loud acoustic signal, either to increase its range of transmission or its ratio to background noise.

Chen [8] studied talker-stress-induced intraword variability and an algorithm that pays off for the systematic changes observed based on hidden Markov models (HMMs) trained by speech tokens under different talking conditions. Raja and Dandapat [9] studied speaker recognition under stressed conditions to improve the decreased performance under such conditions. They used four distinct stressed conditions of SUSAS database. These conditions are neutral, angry, Lombard, and question. They concluded that the least speaker identification performance happened when speakers talk in angry talking environments [9]. Angry talking environments are used as alternatives to shouted talking environments since they can not be totally separated from shouted talking environments in our genuine life [11], [12], [13], [14]. Zhang and Hansen [10] reported on the analysis of characteristics of the speech in five different vocal modes: whispered, soft, neutral, loud, and shouted;

and to recognize discriminating features of speech modes. Shahin focused in four of his earlier studies [11], [12], [13], [14] on improving speaker identification performance in shouted talking environments using each of Second-Order Hidden Markov Models (HMM2s) [11], Second-Order Circular Hidden Markov Models (CHMM2s) [12], Suprasegmental Hidden Markov Models (SPHMMs) [13], and Second-Order Circular Suprasegmental Hidden Markov Models (CSPHMM2s) [14]. The attained speaker identification performance in such talking environments is 59.0%, 72.0%, 75.0%, and 83.4% based on HMM2s, CHMM2s, SPHMMs, and CSPHMM2s, respectively [11], [12], [13], [14].

Most of the works carried out in speech recognition field and speaker recognition field based on HMMs have been conducted using First-Order Hidden Markov Models (HMM1s) [8], [15], [16]. HMM1s give extremely high speaker recognition performance in neutral talking environments [8], [11], [14], while they yield very low performance in shouted talking environments [8], [11], [14]. Mari *et al.* [17], [18] proposed, applied, and tested HMM2s in the training and testing phases of a connected word recognition system under neutral talking condition. They attained very high performance using such models. Shahin [11] exploited these models in the training and testing phases of isolated-word text-dependent speaker identification systems under each of neutral and shouted talking conditions. Based on his work and using HMM2s, Shahin [11] achieved higher speaker identification performance than that using HMM1s under shouted talking condition.

The aim of this work is to propose, implement, and evaluate novel models called Third-Order Hidden Markov Models (HMM3s) to further enhance (compared to HMM2s) text-independent speaker identification performance in shouted talking environments. Speaker recognition in shouted talking environments can be used in criminal investigations to recognize the suspected persons who uttered voice in shouted talking envieonments and in the applications of talking condition recognition. Talking condition recognition can be used in medical applications, telecommunications, law enforcement, and military applications [19]. The proposed models have been evaluated on our collected speech database and SUSAS database.

The rest of the paper is structured as follows: Brief overview of hidden Markov models is given in Section III. The details of the proposed third-order hidden Markov models are covered in Section IV. Section V describes the collected speech database used in this work and the extraction of features. Speaker identification algorithm based on HMM3s and the experiments are discussed in Section VI. Section VII demonstrates the results achieved in the current work and their discussion. Finally, concluding remarks are presented in Section VIII.

## III. BRIEF OVERVIEW OF HIDDEN MARKOV MODELS

HMMs can be described as being in one of the $N$ different states: 1, 2, 3,…, $N$, at any discrete time instant $t$. The individual states are denoted as,

$$s = \{s_1, s_2, s_3, ..., s_N\}$$

which are generators of a state sequence $q_t$, where at any time $t$: q = {q$_1$,q$_2$,…, q$_T$}, $T$ is the length or duration of an observation sequence $O$ of a speech signal. At any discrete time $t$, the model is in a state $q_t$. At the discrete time $t$, the model makes a random transition to a state $q_{t+1}$. The state transition probability matrix $A$ determines the probability of the next transition between states,

$$A = [a_{ij}] \qquad i, j = 1, 2, ..., N$$

where $a_{ij}$ denotes the transition probability from a state $i$ to a state $j$.

## IV. THIRD ORDER HIDDEN MARKOV MODELS

In HMM1s, the underlying state sequence is a first-order Markov chain where the stochastic process is expressed by a 2-D matrix of a priori transition probabilities ($a_{ij}$) between states $s_i$ and $s_j$ where $a_{ij}$ is given as [15], [16],

$$a_{ij} = \text{Prob}\left(q_t = s_j \mid q_{t-1} = s_i\right) \qquad (1)$$

In HMM2s, the underlying state sequence is a second-order Markov chain where the stochastic process is defined by a 3-D matrix ($a_{ijk}$). Therefore, the transition probabilities in HMM2s are given as [17],

$$a_{ijk} = \text{Prob}\left(q_t = s_k \mid q_{t-1} = s_j, q_{t-2} = s_i\right) \qquad (2)$$

with the constraints,

$$\sum_{k=1}^{N} a_{ijk} = 1 \qquad N \geq i, j \geq 1$$

In HMM3s, the underlying state sequence is a third-order Markov chain where the stochastic process is specified by a 4-D matrix ($a_{ijkw}$). Consequently, the transition probabilities in HMM3s are given as,

$$a_{ijkw} = \text{Prob}\left(q_t = s_w \mid q_{t-1} = s_k, q_{t-2} = s_j, q_{t-3} = s_i\right) \qquad (3)$$

with the constraints,

$$\sum_{w=1}^{N} a_{ijkw} = 1 \qquad N \geq i, j, k \geq 1$$

The probability of the state sequence, $Q \triangleq q_1, q_2, ..., q_T$, is defined as:

$$\text{Prob}(Q) = \Psi_{q_1} a_{q_1 q_2 q_3} \prod_{t=4}^{T} a_{q_{t-3} q_{t-2} q_{t-1} q_t} \qquad (4)$$

where $\Psi_i$ is the probability of a state $s_i$ at time $t = 1$, $a_{ijk}$ is the probability of transition from a state $s_i$ to a state $s_k$ at time $t = 3$. $a_{ijk}$ can be computed from (2).

Given a sequence of observed vectors, $O \triangleq O_1, O_2, ..., O_T$, the joint state-output probability is defined as:

$$\text{Prob}(Q, O | \lambda) = \Psi_{q_1} b_{q_1}(O_1) a_{q_1 q_2 q_3} b_{q_3}(O_3) \cdot \prod_{t=4}^{T} a_{q_{t-3} q_{t-2} q_{t-1} q_t} b_{q_t}(O_t) \quad (5)$$

*Extended Viterbi and Baum-Welch Algorithms:*

Based on the probability of the partial alignment ending at a transition $(s_k, s_w)$ at times $(t-1, t)$, the most likely state sequence can be found as:

$$\delta_t(j, k, w) \triangleq \text{Prob}(q_1, \ldots, q_{t-2} = s_j, q_{t-1} = s_k, q_t = s_w, O_1, O_2, \ldots, O_t | \lambda) \quad (6)$$
$$T \geq t \geq 3, N \geq j, k, w \geq 1$$

Recursive computation is given by:

$$\delta_t(j, k, w) = \max_{N \geq i \geq 1} \{\delta_{t-1}(i, j, k) \cdot a_{ijkw}\} \cdot b_w(O_t) \quad (7)$$
$$T \geq t \geq 4, N \geq j, k, w \geq 1$$

The forward function $\alpha_t(j, k, w)$ defines the probability of the partial observation sequence, $O_1, O_2, \ldots, O_t$, and the transition $(s_j, s_k, s_w)$ among times: $t-2$, $t-1$, and $t$ is defined as:

$$\alpha_t(j, k, w) \triangleq \text{Prob}(O_1, \ldots, O_t, q_{t-2} = s_j, q_{t-1} = s_k, q_t = s_w | \lambda) \quad (8)$$
$$T \geq t \geq 3, N \geq j, k, w \geq 1$$

$\alpha_t(j, k, w)$ can be computed from the two transitions: $(s_i, s_j, s_k)$ and $(s_j, s_k, s_w)$ between states $s_i$ and $s_w$ as:

$$\alpha_{t+1}(j, k, w) = \sum_{i=1}^{N} \alpha_t(i, j, k) \cdot a_{ijkw} \cdot b_w(O_{t+1}) \quad (9)$$
$$T - 1 \geq t \geq 3, N \geq j, k, w \geq 1$$

The backward function $\beta_t(i, j, k)$ can be defined as:

$$\beta_t(i, j, k) \triangleq \text{Prob}(O_{t+1}, \ldots, O_T | q_{t-2} = s_i, q_{t-1} = s_j, q_t = s_k, \lambda) \quad (10)$$
$$T - 1 \geq t \geq 3, N \geq i, j, k \geq 1$$

The last equation defines $\beta_t(i, j, k)$ as the probability of the partial observation sequence from $t+1$ to $T$ given the model $\lambda$ and the transition $(s_i, s_j, s_k)$ among times: $t-2$, $t-1$, and $t$.

## V. SPEECH DATABASE AND EXTRACTION OF FEATURES

### A. Collected Speech Database

In the current work, the proposed models have been assessed on our collected speech database. Eight sentences were captured in each of neutral and shouted talking environments in this database. The eight sentences are:

1) He works five days a week.
2) The sun is shining.
3) The weather is fair.
4) The students study hard.
5) Assistant professors are looking for promotion.
6) University of Sharjah.
7) Electrical and Computer Engineering Department.
8) He has two sons and two daughters.

Forty (twenty male students and twenty female students) healthy adult native speakers of American English were asked to utter these sentences. The forty speakers were untrained to avoid exaggerated expressions. Each speaker was separately asked to utter each sentence several times in each of neutral and shouted talking environments. The total number of utterances recorded in both talking environments was 4320 ((40 speakers × first 4 sentences × 9 repetitions/sentence in neutral talking environment) + (40 speakers × last 4 sentences × 9 repetitions/sentence × 2 talking environments)). The collected database was captured in a clean environment by a speech acquisition board using a 16-bit linear coding A/D converter and sampled at a sampling rate of 16 kHz. The database was a wideband 16-bit per sample linear data.

### B. Extraction of Features

In this work, the features that have been adopted to model the phonetic content of speech signals are called Mel-Frequency Cepstral Coefficients (static MFCCs) and delta Mel-Frequency Cepstral Coefficients (delta MFCCs). These coefficients have been used in stressful speech and speaker recognition areas since such coefficients outperform other features in the two areas and because they provide a high-level approximation of human auditory perception [20], [21].

In this work, a 32-dimension feature analysis of both static MFCC and delta MFCC (16 static MFCCs and 16 delta MFCCs) was used to form the observation vectors in each of HMM1s, HMM2s, and HMM3s. The number of states that was used in the experiments was 6 in each model. The number of mixture components, *M*, was 5 per state, with a continuous mixture observation density was selected for each model.

## VI. SPEAKER IDENTIFICATION ALGORITHM BASED ON EACH OF HMM1S, HMM2S, AND HMM3S AND THE EXPERIMENTS

In the training phase of each of HMM1s, HMM2s, and HMM3s (completely three separate phases) the $v^{th}$ speaker model has been derived using the first four sentences of the speech database with 9 repetitions per sentence uttered in the neutral talking environment. The total number of utterances that has been used to derive the $v^{th}$ speaker model in each training phase is 36 (4 sentences × 9 repetitions/sentence). Training of models in HMM1s, HMM2s, and HMM3s training phases uses first-order, second-order, and third-order forward-backward algorithm, respectively.

In the identification phase of each of HMM1s, HMM2s, and HMM3s (completely three separate phases), each one of the forty speakers used separately the last four sentences of the database (text-independent) with 9 repetitions per sentence in each of neutral and shouted talking environments. The total number of utterances that has been used in each phase per talking environment was 1440 (40 speakers × 4 sentences × 9 repetitions/sentence). The probability of generating every utterance per speaker was separately computed based on each of HMM1s,

HMM2s, and HMM3s using Viterbi decoding algorithm. For each one of these three models, the model with the highest probability was chosen as the output of speaker identification as given in the following formula,

a. In HMM1s,

$$V^* = \arg\max_{40 \geq v \geq 1} \left\{ P\left(O \mid \lambda^v_{HMM1s}\right) \right\} \quad (11)$$

where $O$ is the observation vector or sequence that belongs to the unknown speaker and $\lambda^v_{HMM1s}$ is the acoustic first-order hidden Markov model of the $v^{th}$ speaker.

b. In HMM2s,

$$V^* = \arg\max_{40 \geq v \geq 1} \left\{ P\left(O \mid \lambda^v_{HMM2s}\right) \right\} \quad (12)$$

where $\lambda^v_{HMM2s}$ is the acoustic second-order hidden Markov model of the $v^{th}$ speaker.

c. In HMM3s,

$$V^* = \arg\max_{40 \geq v \geq 1} \left\{ P\left(O \mid \lambda^v_{HMM3s}\right) \right\} \quad (13)$$

where $\lambda^v_{HMM3s}$ is the acoustic third-order hidden Markov model of the $v^{th}$ speaker.

## VII. RESULTS AND DISCUSSION

In this work, new proposed models called HMM3s have been employed as classifiers in each of neutral and shouted talking enviroments. These classifiers have been tested on our collected speech databae. Table I summarizes speaker identification performance in neutral and shouted talking environments using the collected database based on each of HMM3s, HMM2s, and HMM1s. This table evidently shows that speaker identification performance in neutral talking environments has been insignificantly improved based on HMM3s compared to that based on each of HMM2s and HMM1s. In neutral talking environments, the average improvement rate of speaker identification performance based on HMM3s compared to that based on HMM2s and HMM1s is 1.6% and 3.3%, respectively. On the other hand, the table apparently illustrates that the performance in shouted talking environments has been significantly enhanced based on HMM3s compared to that based on each of HMM2s and HMM1s. The average improvement rate of speaker identification performance in shouted talking environments based on HMM3s compared to that based on HMM2s and HMM1s is 11.3% and 166.7%, respectively.

A statistical significance test has been performed to show whether speaker identification performance differences (speaker identification performance based on HMM3s and that based on each of HMM2s and HMM1s) are real or simply due to statistical fluctuations. The statistical significance test has been carried out based on the Student's $t$ Distribution test as given by the following formula,

$$t_{1,2} = \frac{\bar{x}_1 - \bar{x}_2}{SD_{pooled}} \quad (14)$$

where $\bar{x}_1$ is the mean of the first sample of size $n$, $\bar{x}_2$ is the mean of the second sample of the same size, and $SD_{pooled}$ is the pooled standard deviation of the two samples given as,

$$SD_{pooled} = \sqrt{\frac{SD_1^2 + SD_2^2}{n}} \quad (15)$$

where $SD_1$ is the standard deviation of the first sample of size $n$ and $SD_2$ is the standard deviation of the second sample of the same size.

Based on Table I and the last two equations, the calculated $t$ values between HMM3s and each of HMM2s and HMM1s in neutral and shouted talking environments using the collected database are given in Table II. Each calculated $t$ value in the neutral talking environments is smaller than the tabulated critical value $t_{0.05} = 1.645$ at *0.05* significant level, while each calculated $t$ value in the shouted talking environments is greater than $t_{0.05} = 1.645$. Therefore, the conclusion that can be drawn in this experiment is that HMM3s insignificantly improve speaker identification performance in neutral talking environments compared to each of HMM2s and HMM1s. It can also be concluded in this experiment that speaker identification performance in shouted talking environments based on HMM3s outperforms that based on each of HMM2s and HMM1s. This significant enhancement in shouted talking environments may be attributed to the fact that in HMM3s the state-transition probability at time $t+1$ depends on the states of the Markov chain at times $t$, $t$-1, and $t$-2. Therefore, the underlying state sequence in HMM3s is a third-order Markov chain where the stochastic process is specified by a 4-D matrix. On the other hand, in HMM2s, the state-transition probability at time $t+1$ depends on the states of the Markov chain at times $t$ and $t$-1. Therefore, the underlying state sequence in HMM2s is a second-order Markov chain where the stochastic process is defined by a 3-D matrix. In HMM1s, it is assumed that the state-transition probability at time $t+1$ depends only on the state of the Markov chain at time $t$. Therefore, in HMM1s the underlying state sequence is a first-order Markov chain where the stochastic process is expressed by a 2-D matrix. Hence, the stochastic process that is specified by a 4-D matrix gives higher speaker recognition performance than that specified by either a 3-D matrix or a 2-D matrix.

In this work, the achieved speaker identification performance based on HMM3s in each of neutral and shouted talking environments is higher than that reported in previous studies [9], [11]. Raja and Dandapat [9] attained 28.57% as an average speaker identification performance in angry talking environments of SUSAS database. Shahin [11] reported an average speaker identification performance of 59.0% in shouted talking environments (collected database) based on HMM2s.

Two extensive experiments have been carried out in this work to evaluate the achieved results of speaker identification performance in each of neutral and shouted talking environments based on HMM3s. The two experiments are:

1. Experiment 1: HMM3s have been assessed on the SUSAS database. This database does not contain shouted talking condition. Since shouted talking condition can not be entirely separated from angry talking condition in real life, HMM3s have been used as classifiers to evaluate speaker identification in angry talking environments. In this experiment, only neutral and angry talking conditions of SUSAS database have been used to assess HMM3s. Table III summarizes speaker identification performance based on each of HMM3s, HMM2s, and HMM1s in neutral and angry talking conditions using such database. The results of this experiment show that HMM3s are superior to each of HMM2s and HMM1s in significantly improving speaker identification performance in angry talking condition. Table IV demonstrates calculated $t$ values between HMM3s and each of HMM2s and HMM1s in the two talking conditions using this database. Table IV evidently shows that HMM3s significantly enhance speaker identification performance compared to each of HMM2s and HMM1s in angry talking condition, while HMM3s insignificantly improve the performance compared to each of HMM2s and HMM1s in neutral talking condition using this database.

2. Experiment 2: An informal subjective assessment of HMM3s using the collected speech database has been performed with ten nonprofessional listeners (human judges). A total of 640 utterances (40 speakers × 2 talking environments × 8 sentences) have been used in this assessment. During this evaluation, each listener was separately asked to identify the unknown speaker in each of neutral and shouted talking environments for every test utterance. The average speaker identification performance in neutral and shouted talking environments based on the subjective assessment is 93.4% and 77.1%, respectively. These averages are close to the averages obtained in the present work using the same database based on HMM3s.

## VIII. CONCLUDING REMARKS

In this work, HMM3s have been proposed, implemented, and evaluated in each of neutral and shouted/angry talking environments as classifiers. Some experiments have been performed to assess these classifiers in the two talking environments. The proposed classifiers have been tested on two distinct speech databases: our collected database and SUSAS database. HMM3s have proven to be superior to each of HMM2s and HMM1s for speaker identification in shouted/angry talking environments, while the proposed models perform slightly better than each of HMM2s and HMM1s in neutral talking environments.

There are some limitations in this work. First, a naïve implementation of the recursion for the computations of $\alpha$ and $\beta$ in HMM3s necessitates on the order of $N^4T$ ($N$ is the number of states and $T$ is the utterance length) operations, compared to $N^3T$ and $N^2T$ operations in HMM2s and HMM1s, respectively. Second, the number of speakers available in SUSAS database is limited to 9. Third, all the 9 speakers available in SUSAS database are of the same gender (male). Finally, speaker identification performance in shouted/angry talking environments based on HMM3s is imperfect.

## REFERENCES

[1] S. Furui, "Speaker-dependent-feature-extraction, recognition and processing techniques," Speech Communication, Vol. 10, March 1991, pp. 505-520.

[2] K. R. Farrell, R. J. Mammone and K. T. Assaleh, "Speaker recognition using neural networks and conventional classifiers," IEEE Transactions on Speech and Audio Processing, Vol. 2, January 1994, pp. 194-205.

[3] K. Yu, J. Mason and J. Oglesby, "Speaker recognition using hidden Markov models, dynamic time warping and vector quantization," IEE Proceedings on Vision, Image and Signal Processing, Vol. 142, No. 5, October 1995, pp. 313-318.

[4] D. A. Reynolds, "Automatic speaker recognition using Gaussian mixture speaker models," The Lincoln Laboratory Journal, Vol. 8, No. 2, 1995, pp. 173-192.

[5] K. E. Cummings and M. A. Clements, "Analysis of the glottal excitation of emotionally styled and stressed speech," Journal of the Acoustical Society of America, Vol. 98, No. 1, July 1995, pp. 88-98.

[6] S. E. Bou-Ghazale and J. H. L. Hansen, "A comparative study of traditional and newly proposed features for recognition of speech under stress," IEEE Transactions on Speech and Audio Processing, Vol. 8, No. 4, July 2000, pp. 429-442.

[7] G. Zhou, J. H. L. Hansen, and J. F. Kaiser, "Nonlinear feature based classification of speech under stress," IEEE Transactions on Speech and Audio Processing, Vol. 9, No. 3, March 2001, pp. 201-216.

[8] Y. Chen, "Cepstral domain talker stress compensation for robust speech recognition," IEEE Transactions on Acoustics, Speech and Signal Processing, Vol. 36, No. 4, April 1988, pp. 433-439.

[9] G. S. Raja and S. Dandapat, "Speaker recognition under stressed condition," International Journal of Speech Technology, Vol. 13, 2010, pp. 141–161.

[10] C. Zhang and J. H. L. Hansen, "Analysis and classification of speech mode: whispered through shouted", INTERSPEECH 2007-ICSLP, 2007, pp. 2289-2292.

[11] I. Shahin, "Improving speaker identification performance under the shouted talking condition using the second-order hidden Markov models," EURASIP Journal on Applied Signal Processing, Vol. 5, issue 4, March 2005, pp. 482-486.

[12] I. Shahin, "Enhancing speaker identification performance under the shouted talking condition using second-order circular hidden Markov models," Speech Communication, Vol. 48, issue 8, August 2006, pp. 1047-1055.

[13] I. Shahin, "Speaker identification in the shouted environment using suprasegmental hidden Markov models," Signal Processing Journal, Vol. 88, issue 11, November 2008, pp. 2700-2708.

[14] I. Shahin, "Employing second-order circular suprasegmental hidden Markov models to enhance speaker identification performance in shouted talking environments," EURASIP Journal on Audio, Speech, and Music Processing, Vol. 2010, Article ID 862138, 10 pages, 2010. doi:10.1155/2010/862138.

[15] J. Dai, "Isolated word recognition using Markov chain models," IEEE Transactions on Speech and Audio Processing Journal, Vol. 3, No. 6, November 1995, pp. 458-463.

[16] L. R. Rabiner, "A tutorial on hidden Markov models and selected applications in speech recognition," Proceedings of IEEE, Vol. 77, No. 2, February 1989, pp. 257-286.

[17] J. F. Mari, J. P. Haton, and A. Kriouile, "Automatic word recognition based on second-order hidden Markov models," IEEE Transactions on Speech and Audio Processing, Vol. 5, No. 1, January 1997, pp. 22-25.

[18] J. F. Mari, F. D. Fohr, and J. C. Junqua, "A second-order HMM for high performance word and phoneme-based continuous speech recognition," Proceedings IEEE International Conference on Acoustics Speech and Signal Processing, Atlanta, USA, Vol. 1, May 1996, pp. 435-438.

[19] J. H. L. Hansen, C. Swail, A. J. South, R. K. Moore, H. Steeneken, E. J. Cupples, T. Anderson, C. R. A. Vloeberghs, I. Trancoso and P. Verlinde, "The impact of speech under stress on military speech technology", NATO Research & Technology Organization RTO-TR-10, Vol. AC/323(IST)TP/ 5IST/TG-01, 2000.

[20] A. B. Kandali, A. Routray, and T. K. Basu, "Emotion recognition from Assamese speeches using MFCC features and GMM classifier," Proc. IEEE Region 10 Conference TENCON 2008, Hyderabad, India, November 2008, pp. 1-5.

[21] T. H. Falk and W. Y. Chan, "Modulation spectral features for robust far-field speaker identification," IEEE Transactions on Audio, Speech and Language Processing, Vol. 18, No. 1, January 2010, pp. 90-100.

Table I

Speaker identification performance in neutral and shouted talking environments using the collected database based on each of HMM3s, HMM2s, and HMM1s

| Models | Gender | Neutral talking environments | Shouted talking environments |
|---|---|---|---|
| HMM3s | Male | 94% | 63% |
| | Female | 95% | 65% |
| | Average | 94.5% | 64% |
| HMM2s | Male | 92% | 57% |
| | Female | 94% | 58% |
| | Average | 93% | 57.5% |
| HMM1s | Male | 92% | 23% |
| | Female | 91% | 25% |
| | Average | 91.5% | 24% |

Table II

Calculated $t$ values between HMM3s and each of HMM2s and HMM1s in neutral and shouted talking environments using the collected database

| Calculated $t$ value ($t_{1,2}$) | Neutral talking environments | Shouted talking environments |
|---|---|---|
| $t_{HMM3s,\ HMM2s}$ | 1.018 | 1.781 |
| $t_{HMM3s,\ HMM1s}$ | 1.345 | 1.822 |

Table III

Speaker identification performance based on each of HMM3s, HMM2s, and HMM1s in neutral and angry talking conditions using SUSAS database

| Models | Neutral talking condition | Angry talking condition |
|---|---|---|
| HMM3s | 95% | 65.5% |
| HMM2s | 93.5% | 58.5% |
| HMM1s | 92% | 27% |

Table IV

Calculated $t$ values between HMM3s and each of HMM2s and HMM1s in neutral and angry talking conditions using SUSAS database

| Calculated $t$ value ($t_{1,2}$) | Neutral talking condition | Angry talking condition |
|---|---|---|
| $t_{HMM3s,\ HMM2s}$ | 1.042 | 1.756 |
| $t_{HMM3s,\ HMM1s}$ | 1.236 | 1.896 |